\documentstyle[12pt,epsfig]{article}
\newcommand\slv{v\kern-5pt\raise1pt\hbox{$\scriptstyle/$}\kern1pt}

\newcommand{\be}{\begin{equation}}
\newcommand{\ee}{\end{equation}}
\def\bq{\begin{eqnarray}}
\def\eq{\end{eqnarray}}
\begin{document}
\thispagestyle{empty}
\begin{flushright}
WUE-ITP-98-043\\
SPhT-T98/019
\end{flushright}
\vspace{0.5cm}
\begin{center}
{\Large \bf Perturbative QCD Correction to the Light-Cone Sum Rule 
for the $B^*B\pi $ and $D^*D\pi$  Couplings}\\[.3cm]
\vspace{1.7cm}
{\sc \bf A.~Khodjamirian$^{1,a} $, R.~R\"uckl$^{1}$, S.~Weinzierl$^{2,b}$, 
O.~Yakovlev$^{1}$}\\[1cm]
\begin{center} \em $^1$ Institut f\"ur Theoretische Physik, 
Universit\"at W\"urzburg,
D-97074 W\"urzburg, Germany \\
\vspace{4mm}
$^2$  Service de Physique Th\'eorique, Centre d'Etudes de Saclay,\\ 
F-91191 Gif-sur-Yvette Cedex, France
\end{center}\end{center}
\vspace{2cm}
\begin{abstract}\noindent
{ The $B^*B\pi$ and $D^*D\pi$ couplings have previously been derived 
from a QCD light-cone sum rule in leading order.
Here, we describe the calculation of the $O(\alpha_s)$ correction 
to the twist 2 term of this sum rule. The result 
is used for a first next-to-leading order analysis. 
We obtain  $g_{B^*B\pi}= 22\pm 7$ and $g_{D^*D\pi}=10.5\pm 3$,
where the error indicates the remaining theoretical uncertainty.
}    

\end{abstract}

\vspace*{\fill}

\noindent $^a${\small \it On leave from 
Yerevan Physics Institute, 375036 Yerevan, Armenia } \\
 $^b${\small \it now 
at NIKHEF, P.O. Box 41882, 1009-DB Amsterdam, The Netherlands}
\newpage

{\bf 1.} The hadronic $B^*B\pi$ 
coupling is defined by the on-shell matrix element
\be
\langle \bar{B}^{*0}(p)~\pi^-(q)\mid B^-(p+q)\rangle =
-g_{B^*B\pi}(q \cdot\epsilon )~,
\label{def}
\ee
where the meson four-momenta are given in brackets and $\epsilon_\mu$ is the 
polarization vector of the $B^*$. An analogous definition holds
for the $D^*D\pi$ coupling. These couplings play an important role
in $B$ and $D$ physics. For example, they  determine the magnitude
of the weak $B\to\pi$  and $D\to\pi$ form factors 
at zero pion recoil, i.e., at  momentum transfer squared close 
to $(m_B-m_\pi)^2$ and $(m_D-m_\pi)^2$, respectively.
Moreover, the coupling  constant $g_{D^*D\pi}$ 
is directly related to the decay width of 
$D^*\to D \pi$. The decay $B^*\to B\pi$ is kinematically forbidden.

Theoretically, the 
$B^*B\pi$ and $D^*D\pi$ couplings have been 
studied using a variety of 
methods\footnote{An overview is given, e.g., in Tab. 1 of ref. \cite{BBKR}.}. 
Among these, QCD light-cone sum rules (LCSR) have proved particularly 
powerful. The first LCSR calculation of 
$g_{B^*B\pi}$ and $g_{D^*D\pi}$ including 
perturbative QCD effects in leading order (LO) 
was reported in \cite{BBKR}. It is well known, however, 
that in heavy-light systems next-to-leading order (NLO) effects
can be essential. 
An important example is provided by the 
heavy-to-light form factors \cite{KRWY,Bagan,BB98,Ball}.
This calls for a NLO analysis of $g_{B^*B\pi}$ and $g_{D^*D\pi}$.

In this letter, we describe the calculation 
of the $O(\alpha_s)$ correction to 
the leading twist 2 term of the relevant LCSR.
The result is then used for a first NLO evaluation. 
We present  numerical predictions for 
$g_{B^*B\pi}$ and $g_{D^*D\pi}$ together with  an estimate 
of the remaining theoretical uncertainties.

\bigskip
{\bf 2. } For definiteness,  we focus 
on the derivation of the $B^*B\pi$ coupling. For 
the  $D^*D\pi$ coupling the procedure is completely analogous. 
The LCSR for $g_{B^*B\pi}$   
is derived from   the vacuum-to-pion correlation function
of the vector and pseudoscalar currents involving the $b$-quark 
and light quark fields: 
$$
F_\mu (p,q)=
i \int d^4xe^{ipx}\langle \pi^-(q)\mid T\{\bar{d}(x)\gamma_\mu b(x),
m_b\bar{b}(0)i\gamma_5 u(0)\}\mid 0\rangle
$$
\be
= F(p^2,(p+q)^2) q_\mu + \tilde{F}(p^2,(p+q)^2) p_\mu~.
\label{corr}
\ee
Inserting in the above matrix element 
complete sets of intermediate hadronic states
carrying $B$ and $B^*$ quantum numbers, respectively,   
and using the definition (\ref{def}) together with 
\be
m_b\langle B^- \mid\bar{b}i\gamma_5 u\mid 0\rangle =m_B^2f_B
\label{fB2}
\ee
and 
\be
\langle 0 \mid \bar{d} \gamma_\mu b \mid \bar{B}^{*0} \rangle = 
m_{B^*} f_{B^*} 
\epsilon_\mu ~,
\label{fBstar2}
\ee 
one obtains double dispersion relations for the invariant functions
$F$ and $\tilde{F}$. In the following, we only need 
\be
F(p^2,(p+q)^2)=\frac{m_B^2m_{B^*}f_Bf_{B^*}g_{B^*B\pi}}{(p^2-m_{B^*}^2)
((p+q)^2-m_B^2)}
+\int\!\!\int\limits_{\hspace{-4mm}\Sigma}
\frac{\rho^h(s_1,s_2)ds_1ds_2}{(s_1-
p^2)(s_2-(p+q)^2)}.
\label{disp}
\ee
The  first term on the r.h.s. is the ground-state contribution 
involving the $B$ and $B^*$ masses, decay constants, and
the desired $B^*B\pi$ coupling.
The second term results from the higher resonances 
and continuum states in the $B^*$ and $B$ channels described by the 
hadronic spectral density $\rho^h(s_1,s_2)$. 
The integration boundary  in the $(s_1,s_2)$-plane is
denoted by $\Sigma$. 
In general, the dispersion relation (\ref{disp}) requires 
subtractions in order to render it finite. However, these 
subtraction terms being 
polynomials in $p^2$ and/or $(p+q)^2$ will be removed by 
Borel transformation\footnote{For the definition of the Borel transformation
see, e.g., ref. \cite{BBKR}.}, and are 
therefore not made explicit here.

In QCD, the invariant amplitude $F(p^2,(p+q)^2)$ 
can be calculated  at $p^2,(p+q)^2\ll m_b^2$ 
by expanding the time-ordered product of currents in the 
correlation function (\ref{corr}) near the light-cone,
that is around $x^2= 0$ (for reviews, see e.g. \cite{KR,Braun}).
The result is given by convolutions of hard scattering amplitudes 
and pion distribution amplitudes. The latter
parameterize the long-distance effects and 
are classified by twist. This paper mainly deals with 
the  leading twist 2 term 
\begin{eqnarray}\label{represent}
F^{(2)}(p^2,(p+q)^2)=-f_\pi\int\limits^1_0 du~ 
\varphi_\pi (u,\mu) T(p^2,(p+q)^2,u,\mu)~,
\end{eqnarray}
$T$ being the hard amplitude,  
$\varphi_\pi$ the twist 2  
distribution amplitude, and $\mu$ the factorization scale.
In analogy to (\ref{disp}), we write 
(\ref{represent}) in the form of a double dispersion relation: 
\be
F^{(2)}(p^2,(p+q)^2)=
\int\!\!\int\frac{\rho(s_1,s_2)ds_1ds_2}{(s_1-
p^2)(s_2-(p+q)^2)}~,
\label{doubled}
\ee
where the corresponding 
spectral density $\rho$ is given by
\begin{eqnarray}
\label{rho12}
\rho(s_1,s_2)= \frac1{\pi^2} \mbox{Im}_{s_1} \mbox{Im}_{s_2}F^{(2)}(s_1,s_2)~.
\end{eqnarray}
After equating (\ref{disp}) and (\ref{doubled}) 
and applying Borel transformation one obtains
the basic sum rule
$$
m_B^2m_{B^*}f_Bf_{B^*}g_{B^*B\pi} 
\exp\left(-\frac{m_{B^*}^2}{M_1^2}-\frac{m_{B}^2}{M_2^2}\right)
+\int\!\!\int\limits_{\hspace{-4mm}\Sigma} \rho^h(s_1,s_2)\exp\left(-\frac{s_1}{M_1^2}
-\frac{s_2}{M_2^2}\right)ds_1ds_2
$$
\bq
=\int\!\! \int 
\rho(s_1,s_2) \exp \left(-\frac{s_1}{M_1^2}-
\frac{s_2}{M_2^2}\right)ds_1 ds_2 
\label{hadr}
\eq
with $M_1^2$ and $M_2^2$  being the Borel parameters associated with
$p^2$ and $(p+q)^2$, respectively.

To proceed, one has to  
subtract from (\ref{hadr}) the unknown contribution of
the excited $B$ and $B^*$ states and of
the continuum states.
This can be achieved approximately by 
using quark-hadron duality. 
To this end, the integral over $\rho^{h}$ is replaced 
by a corresponding integral over $\rho$ with 
a suitable integration boundary $\widetilde{\Sigma}$.
After subtraction, one has 
\bq
f_Bf_{B^*}g_{B^*B\pi}=\frac{1}{m_B^2m_{B^*}}
\int\!\!\int\limits^{\hspace{-2mm}\widetilde{\Sigma}}
\rho(s_1,s_2) \exp \left(\frac{m_{B^*}^2-s_1}{M_1^2}
+\frac{m_B^2-s_2}{M_2^2}\right) 
ds_1 ds_2~.
\label{sr}
\eq
It should be noted that this LCSR actually determines 
the product of the $B$ and $B^*$ decay constants and the $B^*B\pi$
coupling. This will play an important role in the numerical analysis.

\bigskip
{\bf 3.}  The perturbative hard scattering amplitude $T$ introduced in  
(\ref{represent}) can be expanded in a power series in $\alpha_s$:
\begin{eqnarray}
T(p^2,(p+q)^2,u,\mu)=T_0(p^2,(p+q)^2,u)+\frac{\alpha_s C_F}{4\pi}
T_1(p^2,(p+q)^2,u,\mu)+ O(\alpha_s^2)~.
\label{T}
\end{eqnarray}
The lowest-order term is given by \cite{BBKR} 
\begin{equation}
 T_0(p^2,(p+q)^2,u)=-\frac{m_b^2}{m_b^2-p^2(1-u)-(p+q)^2u}~,
\label{Born}
\end{equation}
where $m_b$ is defined to be pole mass of the $b$ quark. 
The $O(\alpha_s)$ term $T_1$ has recently 
been calculated \cite{KRWY,Bagan} in the context 
of the LCSR  determination  of the $B\to \pi$ form factor $f^+(p^2)$.
For brevity, we do not repeat the explicit 
expressions here. They can be found in \cite{KRWY}
together with a detailed discussion 
of the treatment of ultraviolet and collinear divergences.  

The pion distribution amplitude $\varphi_\pi$ 
is associated \cite{BL,exclusive}  
with the leading twist 2 term of the matrix element
\footnote{ Here, the light-cone gauge, $x^\mu A^a_\mu = 0$, is  
adopted for the gluon field $A^a_\mu$.} 
\begin{eqnarray} 
\left\langle \pi(q) \left|
\bar{u}(x) \gamma_{\mu} \gamma_{5}
d(0)\right| 0 \right\rangle =
- i q_{\mu} f_{\pi} \int\limits_{0}^{1} du \varphi_\pi(u,\mu) e^{i u q \cdot x} 
+ ...~,
\label{wave}
\end{eqnarray}
where the ellipses stand for terms of higher twist. 
The scale dependence of $\varphi_\pi(u,\mu)$ follows from a Brodsky-Lepage 
evolution equation. In LO,
the solution is given by 
\begin{equation}
\label{wf}
\varphi_\pi(u,\mu) = \varphi_\pi^{as}(u) 
\left( 1 +\sum_{n=1}^{\infty}a_{2n}(\mu)C^{3/2}_{2n}(2u-1)
\right), 
\end{equation}
where $C^{3/2}_{2n}$
are Gegenbauer polynomials and 
$a_{2n}(\mu)$ are multiplicatively renormalizable 
expansion coefficients. While 
\be
\varphi_\pi^{as}(u)= 6u(1-u)
\label{wfas}
\ee
is the asymptotic distribution amplitude, 
the sum in (\ref{wf}) represents non-asymptotic effects 
with $a_{2n}(\mu)$ vanishing at $\mu\to \infty$.
In NLO \cite{NLO}, the coefficients $a_{2n}$ mix
under renormalization because the Gegenbauer polynomials are no longer 
eigenfunctions of the evolution kernel. 
A complete NLO expression for $\varphi_\pi(u,\mu)$ 
can be found  in \cite{kad}.

The spectral density $\rho(s_1,s_2)$ needed in (\ref{sr}) is obtained 
in LO by substituting the hard amplitude (\ref{Born})
and the distribution amplitude (\ref{wf})
in (\ref{represent}) and taking 
the imaginary parts according to (\ref{rho12}).
In NLO, one has to add the $O(\alpha_s)$ hard scattering 
amplitude $T_1$ as specified in (\ref{T}), and use the solution 
of the  NLO evolution equation for  
$\varphi_\pi$. This yields 
\be
\rho(s_1,s_2)= \rho_0(s_1,s_2)+ \frac{\alpha_s C_F}{4\pi}
\rho_1(s_1,s_2)
\label{rhorho}
\ee  
with 
\bq
\label{rho1}
\rho_{0,1}(s_1,s_2)= - f_{\pi} \frac{1}{\pi^2}
\mbox{Im}_{s_1}\mbox{Im}_{s_2} 
\int\limits_{0}^{1} du ~\varphi_\pi(u,\mu) 
T_{0,1}(s_1,s_2,u,\mu)~.
\eq
As can be seen from (\ref{Born}), $T_0$ only has a single 
pole in $(p+q)^2$. Hence, 
\be
\mbox{Im}_{s_2}T_0(s_1,s_2,u) = - \pi \delta\left(1-\frac{s_1}{m_b^2}(1-u)
-\frac{s_2}{m_b^2}u \right)~,
\ee
making the convolution with $\varphi_\pi(u,\mu )$ 
very simple. After taking the imaginary part in $s_1$, one finds the 
formal expression for $\rho_0$ given in \cite{BBKR}. 
In contrast, 
$T_1$ also has cuts in $(p+q)^2$, in addition to the pole. 
Therefore, Im$_{s_2}T_1$ contains both $\delta$- 
and $\theta$-functions in $s_2 $. In the following, we use the
explicit form given in  \cite{KRWY}. To obtain the $O(\alpha_s)$ 
correction $\rho_1(s_1,s_2)$ 
one has to convolute Im$_{s_2}T_1$ with the   
products $\varphi_\pi^{as}C^{3/2}_{2n}$.
This integration should be carried out analytically 
in order to be able to take the imaginary part in $s_1$. 
In principle, this is possible, by considering the integrals of 
$u^k\mbox{Im}_{s_2}T_1$ for each power $k$ separately.
The calculation  is rather cumbersome, but fortunately not really 
necessary. It turns out that the non-asymptotic effects
in $\rho_1$ are much smaller than the overall theoretical uncertainty 
in the LCSR (\ref{sr}). It is therefore sufficient 
to calculate $\rho_1$ using the asymptotic  pion distribution 
amplitude, unless the theoretical accuracy is increased dramatically. 

With the above justification, we single out the asymptotic part of $\varphi_\pi$ 
given in (\ref{wfas}) and perform the integral 
\be
\int\limits_{0}^{1} du \varphi_\pi^{as}(u) 
\mbox{Im}_{s_2} T_1(s_1,s_2,u,\mu)
\label{int3}
\ee
analytically. After taking the imaginary part in $s_1$ we find 
$$
\rho^{as}_1(s_1,s_2)=
 - f_{\pi} \frac{1}{\pi^2}
 \mbox{Im}_{s_1}\mbox{Im}_{s_2} 
\int\limits_{0}^{1} du \varphi_\pi^{as}(u) 
T_1(s_1,s_2,u,\mu)
$$
$$
=f_\pi\frac{(r+1)^2}{\sigma}\Bigg[\Big(2 \pi^2 
+3 \ln\left(\frac{\sigma}{2}\right) 
\ln\left(1+\frac{\sigma}{2}\right) 
$$
$$
- \frac{3(3 \sigma^3 
+ 22 \sigma^2 + 40 \sigma + 24)}{2(2+\sigma)^3} \ln\left(\frac{\sigma}{2}\right) + 6 \mbox{Li}_2\left(-\frac{\sigma}{2}\right) 
 + \frac{3(\sigma^2 + 12 \sigma +12)}{4(2+\sigma)^2} \Big) 
\delta(1-r) 
$$
\be
-6 \frac{r}{1+r} \left( \frac{\sigma}{1+r+\sigma} 
-2 \ln~r + \ln \left( \frac{1+r(1+\sigma)}{1+r+\sigma} \right) \right)
\frac{d^3}{dr^3} \ln|1-r| \Bigg],
\label{rho1as}
\ee
where 
\bq
\label{vari}
\sigma=\frac{s_1}{m_b^2}+\frac{s_2}{m_b^2}-2, ~~
r=\frac{s_1-m_b^2}{s_2-m_b^2}~,
\eq
and $\mbox{Li}_2(x)=-\int\limits_0^x \frac{dt}t \ln(1-t)$
is the Spence function. It is interesting to note 
that $\rho_1^{as}$ contains  no $\ln\mu$ terms. This is because in $T_1$ the 
coefficient of $\ln \mu$ is given by a convolution of $T_0$ 
with  the LO kernel of the Brodsky-Lepage
evolution equation \cite{KRWY}. It thus has to vanish 
when this kernel is folded with the asymptotic distribution amplitude
$\varphi_\pi^{as}$.

Finally, the resulting spectral density (\ref{rhorho}) 
is to be inserted in the LCSR (\ref{sr}). For  
$M_1^2=M_2^2=2M^2$, the leading term coming from 
$\rho_0$ is simply proportional to $\varphi_\pi(1/2,\mu)$
as shown in \cite{BBKR}. Moreover,
the precise form of the duality boundary $\widetilde{\Sigma}$ 
is irrelevant in this case since only the 
integration limits at $s_1=s_2$ enter.  
The NLO correction arising from $\rho_1$,
on the other hand, depends on the shape 
of the integration boundary $\widetilde{\Sigma}$. We have chosen 
the triangle $s_1+s_2 = 2s_0^B$, $s_0^B$ denoting the 
threshold of excited and continuum states in the $B$-channel. 
This choice allows to 
carry out the integration over the variable $r$ defined in 
(\ref{vari}) analytically. 
The final LCSR reads
\begin{eqnarray}
f_Bf_{B^*}g_{B^*B\pi}=\frac{m_b^2f_\pi}{m_B^2m_{B^*}}
e^{\frac{m_{B}^2+m_{B^*}^2}{2M^2}}
\Bigg[ M^2 \left(e^{-\frac{m_b^2}{M^2}} - 
e^{-\frac{s_0^B}{M^2}}\right)\varphi_\pi(1/2,\mu)
\nonumber
\\
+\frac{\alpha_s C_F}{4 \pi}
\int\limits_{2m_b^2}^{2s_0^B} 
f \left( \frac{s}{m_b^2}-2 \right)
e^{-\frac{s}{2M^2}}ds  
+ F^{(3,4)}(M^2,m_b^2,s_0^B,\mu)\Bigg]~ 
\label{finalsr}
\end{eqnarray}
with
\begin{eqnarray}
f(x) & = & \frac{\pi^2}{4} +
3 \ln\left(\frac{x}{2}\right) \ln\left( 1+\frac{x}{2}\right ) 
- \frac{3(3 x^3 + 22 x^2 + 40 x + 24)}{2(2+x)^3} 
\ln\left(\frac{x}{2}\right) 
\nonumber
\\ 
& & + 6 \mbox{Li}_2\left(-\frac{x}{2}\right) - 3 \mbox{Li}_2(-x) - 
3 \mbox{Li}_2(-x-1) 
-3 \ln(1+x) \ln(2+x) 
\nonumber
\\
& & -\frac{3(3 x^2 + 20 x + 20)}{4(2+x)^2}
+ \frac{6x (1+x) \ln(1+x)}{(2+x)^3}~. 
\end{eqnarray}
In (\ref{finalsr}), we have added  the contributions $F^{(3,4)}$ from 
the pion distribution amplitudes of twist 3 and 4.
The twist 3 contribution is quantitatively  
as important as the twist 2 term, and has to be taken into 
account in phenomenological applications.
Presently, $F^{(3,4)}$ is only known in LO \cite{BBKR}.
The lack of knowledge of the NLO corrections hinders a complete NLO analysis.

The analogous LCSR for $f_Df_{D^*}g_{D^*D\pi}$  
is obtained from (\ref{finalsr}) by formally replacing 
$b$ with $c$, $B$ with $D$, and $B^*$ with $D^*$. 

\bigskip
{\bf 4.} 
The values of the parameters used in the numerical evaluation 
of the LCSR (\ref{finalsr}) and 
the two-point sum rules (2SR)\footnote{In NLO, they are given in 
\cite{fBsr}.} 
for the corresponding decay constants are 
discussed in \cite{BBKR} and \cite{KRWY}. 
A brief recapitulation may therefore suffice here.
For the $b$-quark mass and the corresponding continuum threshold 
we take 
\be\label{interv}
m_b=4.7\pm 0.1\quad\mbox{GeV},\qquad 
s_0^B=35\mp 2\quad\mbox{GeV}^2~.
\ee
Variations of $m_b$ and $s_0^B$ within the allowed intervals 
are correlated by the stability requirement 
of 2SR as indicated by the opposite signs in (\ref{interv}). The 
renormalization scale of $\alpha_s$ is put
equal to the factorization scale $\mu$ which will be fixed later. 
The  running coupling constant is taken in two-loop approximation
with $N_f=4$ and  $\Lambda^{(4)}_{\overline{MS}}=315$ MeV 
corresponding to $\alpha_s(m_Z)=0.118$ \cite{PDG}. 
The Borel parameter in LCSR is constrained to the interval 
$6~ \mbox{GeV}^2<M^2<12$ GeV$^2$, where 
the usual conditions of small ($<10\%$) 
twist 4 corrections and moderate ($<30\%$) 
contributions from heavier states are satisfied. 
In 2SR the allowed range is $2~ \mbox{GeV}^2 <M^2<6$ GeV$^2$.
In the charm case, 
the corresponding parameters are  given by 
\be
m_c=1.3\pm 0.1\quad\mbox{GeV},\qquad 
s_0^D = 6\mp 1\quad\mbox{GeV}^2~,
\ee
and $\Lambda^{(3)}_{\overline{MS}}=380$ MeV. The ranges of the 
Borel parameter are  2 GeV$^2 < M^2< 4$ GeV$^2$ in LCSR
and $1~ \mbox{GeV}^2<M^2<2$ GeV$^2$ in 2SR. 
The meson masses are $m_B=5.279 $ GeV, $m_{B^*}=5.325$ GeV, 
$m_D=1.87$ GeV, $m_{D^*}=2.01 $ GeV, 
and $f_{\pi}=132$ MeV. 
Finally, for the pion distribution amplitude at $u=0.5$ 
we adopt the value  $\varphi_\pi(0.5, \mu)=1.2$ at $\mu= 1$ GeV
determined in \cite{BF} from the LCSR for the pion-nucleon coupling. 
This value 
is also consistent with the distribution amplitude  $\varphi_\pi(u,\mu)$
used in \cite{BBKR,KRWY} to calculate the form factor $f^+$.
The values of the vacuum condensates appearing in 2SR can be found in
\cite{KR}.

\begin{figure}[ht]
\mbox{
\epsfig{file=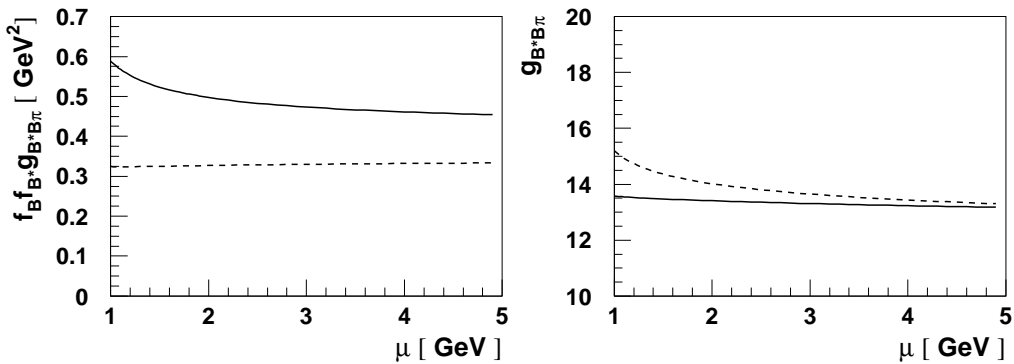,width=\textwidth,bbllx=0pt,bblly=260pt,bburx=600pt,%
bbury=530pt,clip=}
}
\caption{\it Scale dependence of the twist 2 term 
in the sum rule estimates of 
(a) $f_Bf_{B^{*}}g_{B^*B\pi}$  
 and (b) $g_{B^*B\pi}$: NLO (solid) in 
comparison to LO (dashed). }
\end{figure}

Let us first discuss the twist 2 term  in the LCSR (\ref{finalsr})
being the main subject of this paper. This term is plotted in Fig. 1a 
as a function of $\mu$
for $M^2= 8 ~\mbox{GeV}^2$.  
As can be seen, the NLO correction is 
numerically large, increasing $f_Bf_{B^{*}}g_{B^*B\pi} $ by about $50 \%$.
It has the same sign as
the NLO corrections to the 2SR for $f_B$ and $f_B^*$
which amount to about 30 \% and 20\%, respectively. 
Consequently, in the coupling $g_{B^*B\pi}$ itself, the NLO effects cancel 
almost completely as demonstrated in Fig. 1b. 
A similar cancellation takes place in the sum rule calculation
of the $B\to \pi$ form factor $f^+$ \cite{KRWY,Bagan}.
As shown in \cite{Bagan}, the origin of this cancellation 
can be understood
by considering the heavy quark limit of the sum rules. 
In this limit, one has a partial cancellation 
of large $\ln(m_b/\omega_0)$ appearing in $f_Bf^+$ and $f_B$, 
where $\omega_0=(s_0^B-m_b^2)/2m_b$ determines the size
of the duality interval which is mass-independent. 
In the present case, we observe a complete cancellation of these logarithms 
in the heavy-quark limit of the LCSR
(\ref{finalsr}) and the 2SR for $f_Bf_{B^*}$.
As a result, the NLO correction to the coupling constant
$g_{B^*B\pi}$ contains no logarithms of $m_b$ at all. Furthermore, 
the scale-dependence of  $f_Bf_{B^{*}}g_{B^*B\pi}$ 
is moderate. In LO, it only comes from 
the evolution of the distribution amplitude $\varphi_\pi$ , while in NLO it 
also arises from the running of $\alpha_s(\mu)$.
Making the usual choice  
$\mu =\sqrt{m_B^2-m_b^2} = 2.4 ~\mbox{GeV}$, a scale which is
of the order of the Borel mass $M$,  
and allowing $\mu$ to vary from one half to twice of the above value, 
$f_Bf_{B^{*}}g_{B^*B\pi}$ (NLO) 
varies by $10\%$ relative to the nominal value. Again, this is 
comparable to the NLO scale-dependence of $f_B$ and $f_B^{*}$
so that $g_{B^*B\pi}$ is almost scale-independent
as can be seen in Fig. 1b.

Similar features are observed in the $D$-meson case. In particular, 
the NLO correction to $f_Df_{D^{*}}g_{D^*D\pi}$ of roughly 15\% 
is compensated by the NLO corrections to $f_{D}$ and $f_{D^*}$
of about 10\% each. 
It may look surprising that  the perturbative corrections   
are smaller in the charm case than in the beauty case,
in contrast to the naive expectation from asymptotic freedom.
The reason becomes again clear in the heavy quark limit:
when going from beauty to charm, 
the growth of $\alpha_s(\mu)$ is overcompensated by the shrinkage 
of the above-mentioned logarithm $\ln(m_b/\omega_0)$.
 
\begin{table}[t]
\caption{\label{tab1}
{\it Sum rule predictions for 
$B$ and $B^*$ mesons}}
\begin{center}
\begin{tabular}{|c||c|c|}
\hline
&&\\
LCSR & $f_Bf_{B^{*}}g_{B^*B\pi} $ & 0.80$\pm 0.23~~\mbox{\rm GeV}^2$   ~~~\\
&&\\
\hline
&&\\
2SR & $f_B$ & 180$\pm 30~~\mbox{\rm MeV}$   ~~~\\
&&\\
\hline
&&\\
$\frac{\mbox{LCSR}}{\mbox{2SR}}$ & 
$f_{B^{*}}g_{B^*B\pi}$ & 4.44$\pm 0.97~~\mbox{\rm GeV}$   ~~~\\
&&\\
\hline
&&\\
2SR & $f_{B^*}$ & 195$\pm 35~~\mbox{\rm MeV}$   ~~~\\
&&\\
\hline
&&\\
$\frac{\mbox{LCSR}}{\mbox{2SR}}$ & $g_{B^*B\pi} $ &22 $\pm$ 7   ~~~\\
&&\\
\hline
\end{tabular}
\\
\end{center}
\hspace*{1.5cm}
\end{table}

\begin{table}[t]
\caption{\label{tab2}
{\it Sum rule predictions for 
$D$ and $D^*$ mesons}}
\begin{center}
\begin{tabular}{|c||c|c|}
\hline
&&\\
LCSR 
& $f_Df_{D^{*}}g_{D^*D\pi} $ & 0.54$ \pm 0.15 ~~\mbox{\rm GeV}^2$ \\
&&\\
\hline
&&\\
2SR 
& $f_D$ & 190$\pm 20~~\mbox{\rm MeV}$   ~~~\\
&&\\
\hline
&&\\
$\frac{\mbox{LCSR}}{\mbox{2SR}}$ 
& $f_{D^{*}}g_{D^*D\pi} $ &2.84$\pm 0.55~~\mbox{\rm GeV}$   ~~~\\
&&\\
\hline
&&\\
2SR 
& $f_{D^*}$ & 270$\pm 35~~\mbox{\rm MeV}$   ~~~\\
&&\\
\hline
&&\\
$\frac{\mbox{LCSR}}{\mbox{2SR}}$ 
& $g_{D^*D\pi} $ & 10.5 $\pm$ 3   ~~~\\
&&\\
\hline
\end{tabular}
\\
\end{center}
\hspace*{1.5cm}
\end{table}

The numerical predictions including the twist 3 and 4 contributions 
in LO \cite{BBKR} are summarized in Tables 1 and 2.  
In dividing out the pseudoscalar and/or vector meson 
decay constants from the LCSR results, 
one can follow different procedures. The simplest 
possibility is to divide the complete sum rule (\ref{finalsr})   
by the NLO values of these constants. Alternatively, one may 
divide the twist 2 term  in (\ref{finalsr}) by the NLO values of 
decay constants, while dividing the twist 3 and 4 terms
by the LO values. The numerical difference lies within the 
overall theoretical uncertainty quoted in the Tables. 
We prefer the first option.
For $g_{B^*B\pi}$, the uncertainty can be estimated as follows:  

(a) Variation of the Borel parameters in the allowed 
ranges  leads to a variation of the coupling by about 10\%.

(b) If the $b$-quark mass and the continuum threshold $s_0^B$ 
are varied simultaneously such that 
maximum stability of the sum rules for $f_B$
and $f_B^*$ is achieved,  $g_{B^*B\pi}$ varies by 25\%.

(c) The uncertainty coming from 
the non-asymptotic terms in  the pion distribution amplitudes
amounts at most 10\%. This can be checked by 
discarding the non-asymptotic effects altogether.

(d) For the unknown NLO corrections beyond twist 2 we assign a 15\%
uncertainty. Since twist 3 contributes about 50\% to the sum rule 
(\ref{finalsr}) this actually assumes an unknown  
correction to twist 3 and beyond as large as 30\%.

(e) The twist 4 contribution to the LCSR 
(\ref{finalsr}) is roughly 3\%. This can be taken 
as an indication of the uncertainty due to the neglect 
of twists higher than four.

(f) The size of the non-asymptotic terms in $\rho_0$ implies 
that they should affect the 
$O(\alpha_s)$ spectral density $\rho_1$ by less than $5\%$. 

(g) In (\ref{sr}) there is some ambiguity in the  choice of the 
duality boundary $\widetilde{\Sigma}$ for $\rho_1$. 
A reasonable way to estimate the maximum uncertainty 
is to put the upper integration limit $s_0^B$ in (\ref{finalsr}) 
to infinity which corresponds to no continuum 
subtraction at all. The change is less than 1\%. 

(h) The sensitivity to the precise value of the quark condensate density 
is rather small, because of a compensation of the 
effects in 2SR and in the twist 3 term of LCSR. 
The inaccuracy of higher-dimensional condensates in
2SR is unimportant.

Adding (a) to (h) in quadrature, the overall uncertainty on 
$g_{B^*B\pi}$ is shown in Table 1 together with the uncertainties 
in the related quantities estimated in a similar way.
The corresponding error estimates for the $D$-meson case are 
listed in Table 2.

A comprehensive comparison with predictions obtained by different 
methods can be found in Table 1 of ref. \cite{BBKR}. New estimates of 
$g_{B^*B\pi}$ in the 
framework of lattice QCD \cite{lat} and in a relativistic 
quark model \cite{ley} have appeared very recently.
While our result shown in Table 1 agrees with the lattice estimate 
within uncertainties, it is smaller by a factor of 2-3 
than the quark-model prediction.

\bigskip
{\bf 5.} As pointed out in the introduction,
the $B^*B\pi$ matrix element determines 
the $B\to \pi$ form factor $f^+(p^2)$  near the kinematic limit 
$p^2 = (m_B-m_\pi)^2$, where the $B^*$-pole dominates. 
More definitely, in the single-pole
approximation, one has
\be
f^+(p^2)= \frac{f_{B^*}g_{B^*B\pi}}{2m_{B^*}(1-p^2/m_{B^*}^2)}~
\label{onepole}
\ee 
with the normalization $f_{B^*}g_{B^*B\pi}$ given in Table 1.
Since the NLO corrections to the approximation (\ref{onepole}) 
at large $p^2$ and to $f^+(p^2)$ 
calculated from LCSR directly at low and intermediate $p^2$
are both small, the two descriptions
match in NLO as well as they do in LO \cite{BBKR}. 
This also holds for the $D$-meson form factor.   

Furthermore, from $g_{D^*D\pi}$ given in Table 2,
one obtains : 
\be
\Gamma( D^{*+} \rightarrow D^0 \pi^+)~ =
~ \frac{g_{D^*D\pi}^2}{24\pi m_{D^*}^2}|~\vec q ~|^3  
~= ~23 \pm 13~\mbox{keV},
\ee
$\vec q~$  being the decay 3-momentum. 
The current experimental limit \cite{PDG} 
\begin{eqnarray}
\Gamma (D^{*+}\to D^0\pi^+) < 89~\mbox{keV}
\end{eqnarray}
is still too high to challenge the theoretical prediction.

In conclusion, we have made some progress in the NLO 
analysis of the LCSR for the $B^*B\pi$
and  $D^*D\pi$ coupling constants. 
The perturbative $O(\alpha_s)$-correction to 
the twist 2 term of this sum rule 
has been calculated removing one of the main uncertainties.
The remaining uncertainties in 
$g_{B^*B\pi}$ and $g_{D^*D\pi}$ are mainly due to 
the heavy quark masses, the uncertainty 
in the pion distribution amplitudes of 
twist 2 and 3 at $ u \simeq 1/2 $, and the lack of 
the NLO corrections to the twist 3 contributions.
Thus, there is still plenty of room for improvement.

\bigskip

{\bf Acknowledgments}

We are grateful to V.M.~Braun for useful discussions.
This work was supported by the German Federal Ministry for 
Education and Research (BMBF) under contract number 05 7WZ91P (0).

\end{document}